\documentclass[sigconf]{acmart}
\usepackage{amsmath,amssymb,amsfonts}
\usepackage{algorithmic}
\usepackage{graphicx}
\usepackage{textcomp}
\usepackage{xcolor}
\usepackage{algorithm}
\usepackage{multirow}
\usepackage{multicol}

\renewcommand\footnotetextcopyrightpermission[1]{} 
\begin{document}

\title{Dynamic Anonymized Evaluation for Behavioral Continuous Authentication}

 \author{Rasana Manandhar}
 \email{rmanandh@uwyo.edu}
 \affiliation{
   \institution{University of Wyoming}
   \streetaddress{1000 E University Ave}
   \city{Laramie}
   \state{Wyoming}
   \postcode{82071}
 }

 \author{Shaya Wolf}
 \email{swolf4@uwyo.edu}
 \affiliation{
   \institution{University of Wyoming}
   \streetaddress{1000 E University Ave}
   \city{Laramie}
   \state{Wyoming}
   \postcode{82071}
 }

 \author{Dr. Mike Borowczak}
 \email{mike.borowczak@uwyo.edu}
 \affiliation{
   \institution{University of Wyoming}
   \streetaddress{1000 E University Ave}
   \city{Laramie}
   \state{Wyoming}
   \postcode{82071}
 }


\begin{abstract}
Emerging technology demands reliable authentication mechanisms, particularly in interconnected systems. Current systems rely on a single moment of authentication, however continuous authentication systems assess a users identity utilizing a constant biometric analysis. Spy Hunter, a continuous authentication mechanism uses keystroke dynamics to validate users over blocks of data. This easily-incorporated periodic biometric authentication system validates genuine users and detects intruders quickly. Because it verifies users in the background, Spy Hunter is not constrained to a password box. Instead, it is flexible and can be layered with other mechanisms to provide high-level security. Where other continuous authentication techniques rely on scripted typing, Spy Hunter validates over free text in authentic environments. This is accomplished in two phases, one where the user is provided a prompt and another where the user is allowed free access to their computer. Additionally, Spy Hunter focuses on the timing of different keystrokes rather than the specific key being pressed. This allows for anonymous data to authenticate users and avoids holding personal data. Utilizing a couple K-fold cross-validation techniques, Spy Hunter is assessed based on how often the system falsely accepts an intruder, how often the system falsely rejects a genuine user, and the time it takes to validate a users identity. Spy Hunter maintains error rates below 6\% and identifies users in minimal numbers of keystrokes. Continuous authentication provides higher level security than one-time verification processes and Spy Hunter expands on the possibilities for behavioral analysis based on keystroke dynamics. 
\end{abstract}

\maketitle

\section{Introduction}
The permeation of technology throughout our society now dictates a need for determining the authenticity of access to interconnected systems. Current authentications methods rely on instantaneous one-time verification that are easily penetrated or circumvented due to inherent system weaknesses and vulnerabilities or user error.  Generally, systems require some authentication mechanism to verify a user's identity before gaining access. However, these systems often rely on a singular instance of verification, usually based on the use of 'something known,' in order to provide an instantaneous pseudo-guarantee of a user's identity. Interconnected systems that rely on one-time verification are vulnerable to adversaries capable of spoofing passwords or bypassing authentication measures. Conversely, continuous authentication mechanisms constantly verify a user's identity by analyzing how a user completes certain tasks to create a behavioral fingerprint for that user. 

Spy Hunter, a continuous authentication mechanism, uses keystroke time intervals to validate users over blocks of data. By detecting intruders over minimal frames of data, Spy Hunter narrows the attack vector to a minuscule window of opportunity. This system was tested using k-fold cross validation based on two sets of data. Invalid users were detected consistently and quickly without compromising the usability of the device for valid users. 

\section{Applications}
Continuous authentication measures can be applied to existing systems to improve security. Some of the most important systems, namely those that hold confidential information or those that control critical infrastructure, are vulnerable. Medical systems, for instance, hold sensitive and personally identifiable information (PII) such as social security numbers, payment methods, and personal health information. This information forms a target for adversaries and necessitates increased security. Electrocardiogram-based authentication systems  \cite{Zhang_2018} and other verification techniques use signal streams gathered from biomedical devices \cite{Mosenia_2017} to authenticate users. These authentication measures require communication between personal smart devices and the rest of world. Biomedical devices monitor many things about an individual and hold that private data, creating another attack vector. Spy Hunter analyzes a users behavior to validate their identity. Unlike other systems which capture specific user data, Spy Hunter captures only the timing of key presses rather than the specific key a user is typing. 

Emerging technology allows us more connectivity from our mobile devices, creating high demand for continuous authentication approaches. Security mechanisms should be implemented on mobile devices in consideration for the immense amount of personal data they hold. Mobile continuous authentication systems utilize touch biometrics  and keystroke dynamics to verify users accurately. Devices employing touch-based continuous authentication rely on raw touch data and aggregated gesture data \cite{Kroeze_2016}. However, employing keystroke dynamics allow for much more flexibility across platforms since most devices have key press equivalents. Current research using keystroke dynamics yields promising results, but high error rates \cite{Saevanee_2015}. Other mobile continuous authentication systems focus on verifying users quickly, using Bayesian and MiniMax versions of quickest change detection algorithms to provide continuous authentication\cite{Perera_2018}. Although these systems are aimed at quickly detecting invalid users, they are constrained by accuracy requirements. 

Continuous authentications are not constrained to various machines. Some systems utilize physical equipment to verify users. For instance, facial recognition systems require some sort of camera. Additionally, fingertip sensor devices can be used to authenticate based on captured behavioral characteristics\cite{Wu_2017}. Much like the biomedical devices described, these mechanisms necessitate additional hardware to implement. However, ideal security measures place minimal impact on genuine users. Spy Hunter provides promising preliminary results without requiring additional hardware. 

\section{Motivations}
Continuous authentication mechanisms utilize biometrics to provide higher promises of security while remaining easy to incorporate in currently deployed systems. Keystroke dynamics, the typing patterns and rhythms of the user, show particular promise through low-cost integration and portability \cite{Ali_2017}. Because most devices utilize a keyboard, typing dynamics are measurable across different devices, making keystroke evaluation systems like Spy Hunter portable and convenient for continuous authentication. Using keystroke dynamics through different stages of identification and verification provide better, more cost-effective, authentication \cite{Patel_2018}. Novelty search algorithms, using user-adaptive feature extraction methods rather than layout-based feature extraction can validate users as well\cite{Kim_2017}. Spy Hunter however, does not rely on which key was pressed. Because the keys are anonymized, feature extraction methods are complicated. Nevertheless, Spy Hunter allows more flexibility across users and allows for a seamless authentication system for bilingual users. 

Individuals, businesses, industries, and governments rely on valid authentication of genuine users and quick detection of invalid users. Validation frameworks show potential for continuous authentication, motivating their improvement \cite{Abdulwahid_2015}. Specifically, continuous authentication systems can be improved by layering different techniques together. Some continuous systems focus only on the strength of their own individual authentication technique. More adaptable systems create modular schemes that can be combined with other dynamic evaluation techniques to provide multi-faceted authentication. Spy Hunter authenticates against the keystrokes across an entire device, meaning it isn't bound to a password box or a specific program. Background authentication allows freedom to the user as well as providing a modular approach to validation. Subsequently, Spy Hunter can be combined with other security measures to construct a multi-modal authentication system. Multi-modal authentication systems create dynamic trust models and protect against avoidance attacks where an intruder limits their input device usage to skirt around a single continuous authentication mechanism \cite{Mondal_2017}.

Multi-modal authentication systems promote the use of multiple sensors. Some systems use many sensors across different categories, specifically keystroke dynamics, mouse movement, and stylometry to gather data. Using multiple authentication techniques allows false accept rates (FAR) and false reject rates (FRR) to drop below .1\% and .2\% respectively, with mouse curve distance being particularly useful \cite{Fridman_2014}. However, when sensors are compromised, the system degrades quickly. Consequently, multi-modal systems rely on separately secure modular authentication systems rather than one system evaluating multiple behaviors at once. Spy Hunter can be deployed alongside authentication systems that evaluate a users validity through different metrics, such as mouse movements. Continuous biometric authentication methods based on mouse movements can benefit keystroke dynamic analysis \cite{Mondal_2015}, making the pair of metrics particularly cohesive. 

In summary the primary contributions of this work include:
\begin{enumerate}
    \item An easily-incorporated periodic biometric authentication mechanism based on keystroke dynamics
    \item Valid authentication of genuine users and quick detection of invalid users
    \item Flexible authentication that can be layered with other methods to provide accurate verification
    \item Keystroke timing analysis over free text and freedom in testing environment to simulate authentic verification necessities
\end{enumerate}

\section{Methods}
Some keystroke-based continuous authentication systems utilize non-conventional keystroke features like typing speed, error rate, and shift-key usage to garner extra information to achieve higher accuracy rates \cite{Alsultan_2017}. Spy Hunter benefits from a simpler approach that focuses on the timing information behind how all keys are pressed without knowledge of the individual key that was pressed. Methods for evaluating continuous authentication systems based on performance reporting \cite{Bours_2015} dictate that continuous authentication systems act on each single action, while a periodic system acts on a block of actions at a time. Spy Hunter currently evaluates small blocks of data for preliminary evidence of the mechanisms success. By considering blocks of keystrokes at a time, Spy Hunter is able to provide initial success and encourages a sliding-frame technique that provides a genuine continuous authentication mechanism as well as building in redundancy for security purposes.

Spy Hunter first identified a user based on their keystrokes by extracting timing metrics while they type on their machine. Extracting key press and release times was handled through a Python library ('pynput') which runs a thread to capture keys being typed by a user. The library captured the time of each keypress and key release, allowing the timing between consecutive keys to be extracted. Then, features were extracted from the timing between keys. Four features describe a persons typing pattern:
\begin{enumerate}
\item{\textbf{Hold time}: The time duration between the press and release of a single key. }
\item{\textbf{Up Down Key (UD)}: The time duration between the release of one key and press of the next key. }
\item{\textbf{Down Down Key (DD)}: The time duration between the press of one key and the press of the next key.}
\item{\textbf{Up Up Key (UU)}: The time duration between the releases of one key and the release of the next key.}
\end{enumerate}

A One Class Support Vector Machine (SVM) model was selected and trained with features extracted from the genuine user. This model was then used to detect intruders. One Class SVMs find a linear boundary to maximally separate instances of different classes. They are trained with features of one class to create a model which is then used to classify future data. When features are detected that are consistent with the model, they are classified with a label of 1. Any data that differs from the trained data are classified with a label of -1. Spy Hunter uses this to train a model with the features of a genuine user. When the model detects the genuine user's features, those features are classified with a 1, meaning they belong to genuine user. Any typing features that differ enough (based on accuracy constraints) from the genuine user's model are classified as -1, meaning they belong to an intruder. 

\section{Experiment Parameters}
Data was gathered in two experimentation phases. First, users were asked to type according to a prompt to determine authentication viability when users are in similar mindsets. Then, after 2,000 characters were gathered from the prompt, users were given free use of their computers while 2,000 more characters were gathered in the background. This measured authentication viability when users were given freedom to act normally. The first data collection phase simulates validation allowing initial access to a machine. The second phase mirrors continuous background authentication while the user is going about their normal business. 

After data collection, a preliminary and advanced evaluation of Spy Hunter yielded similar results. The preliminary evaluation of the Spy Hunter mechanism utilized a multiple K-Fold approach. Assessing each phase separately determined which data collection technique garnered more promising results. Our initial K-Fold approach segmented each 2,000 character data set into a training and testing set, using 1,500 characters for training and the remaining 500 characters for testing. The test data was broken down further into different size blocks of data to gauge how many characters were required to authenticate a genuine user or reject an intruder. Block sizes of 30, 50, 80, and 100 were considered. 

Following the preliminary evaluation, Spy Hunter was evaluated again forming the training and testing data sets randomly. Data was first split into 5 folds, then 10 of equal size. One fold was randomly selected to test the model, leaving the rest of the data for training. The training data established a model for each genuine user, characterizing their normal behavior. These models were evaluated across every other users testing data to determine the authenticity of each user. Users were verified based on blocks of test data, each holding timing logistics for 80 characters. 

During verification, if the testing data and training data belong to two different individuals, the identity of the user is predicted based on the testing data, in conjunction with a specified level of certainty. The best results occurred with a 65\% accuracy threshold. The accuracy of the prediction was assessed against this threshold to determine if the user should be allowed or rejected. If the accuracy of the prediction is above the certainty threshold, the user is rejected, otherwise the user is allowed continued access to the system. This continues, grabbing a block of data and assessing the accuracy of the user's identity prediction until the user is rejected. If the user is not rejected before all of the data is exhausted, they are considered to be falsely accepted. 

\begin{figure}
\includegraphics[width=\linewidth]{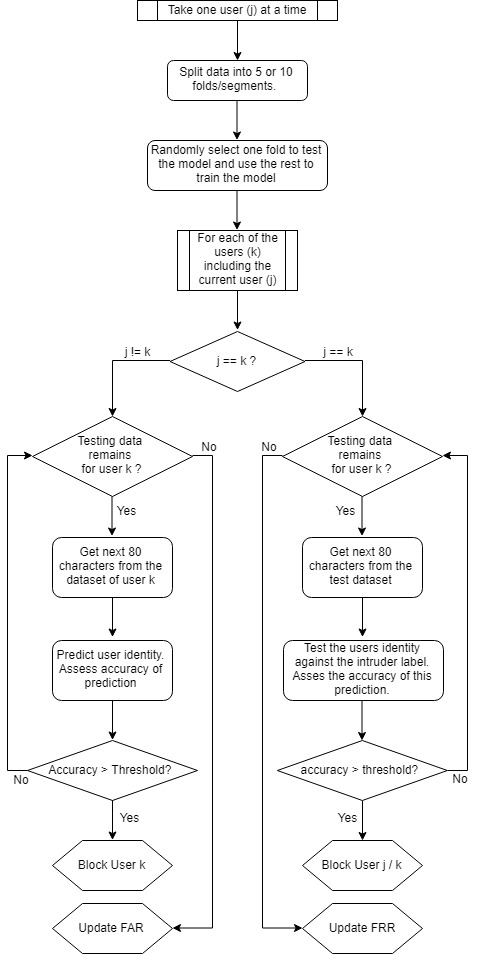}
\caption{ Extended experiment process }
\label{flowchart}
\end{figure}

If the testing data and training data belong to the same individual, the identity of the user is still predicted based on the testing data, but their identity is assessed based off of the intruder label. Instead of assuming the user is authentic and seeing if Spy Hunter can detect them as invalid, they are assumed an intruder to see if Spy Hunter can detect them as actually valid. Again, they are tested against a threshold of certainty. If Spy Hunter determines, with greater than 65\% accuracy, that the user is invalid, the genuine user will be inaccurately blocked from the system. Conversely, if the accuracy is below 65\%, the system allows the genuine user to continue on the system. This continues until all of the training data is exhausted. If Spy Hunter fails to authenticate a valid user against an intruder label, they are considered to be falsely rejected.  This process can be seen in Figure \ref{flowchart}.

\section{Results}
The initial effectiveness of Spy Hunter was evaluated using three metrics:
\begin{enumerate}
\item{\textbf{FAR}: The False Accept Rate, or the percentage of times Spy Hunter detects an intruder as a genuine user.}
\item{\textbf{FRR}: The False Reject Rate, or the percentage of times Spy Hunter detects a genuine user as an intruder.}
\item{\textbf{Number of Blocks}: The average number of blocks it takes for Spy Hunter to detect a user as genuine or an intruder.}
\end{enumerate}

The initial experimentation showed surprisingly exceptional results for the 20 participants. For the first phase of data, where users typed in a specified window given a prompt, false acceptance rates (FAR) and false rejection rates (FRR) were held below 6\% in all cases. The highest error rate is the FAR for block sizes of 100 characters. Under this configuration, Spy Hunter accepts an intruder 5.25\% of the time. However, it only rejects a genuine user 1.25\% of the time. Additionally, Spy Hunter required limited data to make validity decisions, keeping the average number of blocks of data between two and three blocks. These block size variations show that the amount of data in a block does not indicate the system performance. When bigger blocks are employed, intruders are more likely to gain access, but when smaller blocks are used, genuine users are more likely to be rejected. These results can be seen in Table \ref{table1}. 

\begin{table}[H]
\begin{tabular}{l|c|c|c|c|}
\cline{2-5}
\multirow{2}{*}{}                       & \multicolumn{4}{l|}{Block Size (\# Characters per Block)} \\ \cline{2-5} 
                                        & 30           & 50           & 80           & 100          \\ \hline
\multicolumn{1}{|r|}{FAR}               & 0.0025       & 0.0150       & 0.0400       & 0.0525       \\
\multicolumn{1}{|r|}{FRR}               & 0.0500       & 0.0350       & 0.0175       & 0.0125       \\
\multicolumn{1}{|l|}{Avg. \# of Blocks} & 2.1947       & 2.6684       & 3.0711       & 2.9737       \\ \hline
\end{tabular}
\caption{Results for first phase of initial experimentation}
\label{table1}
\end{table}

For the second phase of data, where users had free use of their machine, Spy Hunter authenticated users with a 0\% FAR in some cases. This means that with a block size of 30 or 50, an intruder never gained control over the system. Smaller block sizes in this experiment resulted in lower FARs, but higher FRRs, depicting the trade off between having higher/lower accuracy thresholds. For example, even though the FAR was 0\% for block sizes of 30 and 50, the FRR was 5\% and 4\% respectively. Further, the average number of blocks required to determine a users authenticity decreased between the first and second phase of data analysis. The best performing block size in the first phase contained 30 characters and used 2.1947 blocks on average to authenticate a user. In contrast, a block size of 30 characters used 1.8237 blocks to authenticate in the second phase. These results can be seen in Table \ref{table2}.

\begin{table}[H]
\begin{tabular}{l|c|c|c|c|}
\cline{2-5}
\multirow{2}{*}{}                       & \multicolumn{4}{l|}{Block Size (\# Characters per Block)} \\ \cline{2-5} 
                                        & 30           & 50           & 80           & 100          \\ \hline
\multicolumn{1}{|r|}{FAR}               & 0.0000       & 0.0000       & 0.0100       & 0.0150       \\
\multicolumn{1}{|r|}{FRR}               & 0.0500       & 0.0400       & 0.0300       & 0.0225       \\
\multicolumn{1}{|l|}{Avg. \# of Blocks} & 1.8237       & 2.2763       & 2.4553       & 2.5737       \\ \hline
\end{tabular}
\caption{Results for second phase of initial experimentation}
\label{table2}
\end{table}

Given these promising results, Spy Hunter shows potential for a continuous authentication system. To further solidify this evidence, Spy Hunter was a evaluated again using random selection of data segments to train and test the data. This second analysis was also split into two phases. Recall that the first 2000 characters were gathered from the prompt and were tested separately from the 2000 characters that were gathered while the user had free access to their computer. Each of these data sets are further split into different data segments and were tested using 5 segments and 10 segments. The segments are chosen randomly and the first selected segment is used as the training segment. Then, each user was compared with every other user. Consistency was maintained in the trade off between FARs and FRRs. Although this extended experimentation required more blocks of data than the initial experimentation, Spy Hunter was still able to keep the average number of blocks below 3. With 80 characters per block, this allows an attack vector of 240 character. The results further solidified the previous results and reduced the likelihood of having over-fit the data during initial experimentation. These results can be seen in Table \ref{table3}.

\begin{table}[H]
\begin{tabular}{c|c||c|c|c|}
\cline{2-5}
\multicolumn{1}{l|}{}                               & \# Folds & FAR    & FRR    & Avg. \# of Blocks \\ \hline
\multicolumn{1}{|c|}{\multirow{2}{*}{First Phase}}  & 5        & 0.0310 & 0.0135 & 2.7932            \\
\multicolumn{1}{|c|}{}                              & 10       & 0.0320 & 0.0068 & 2.9834            \\ \hline
\multicolumn{1}{|c|}{\multirow{2}{*}{Second Phase}} & 5        & 0.0083 & 0.0170 & 2.6813            \\
\multicolumn{1}{|c|}{}                              & 10       & 0.0120 & 0.0115 & 2.8542            \\ \hline
\end{tabular}
\caption{Results for extended experimentation}
\label{table3}
\end{table}

\section{Discussion}
\subsection{Result Implications}
Different block sizes were tested to find an ideal authentication window. Having a small block size limits the attack vector but compromises the predictive ability of the system whereas having a large block size creates a larger authentication window and gives an intruder ample time to make serious changes in the system. For the first phase of the experiment, a block size of 50 seems preferable because it minimizes all three metrics (FAR, FRR, and Number of Blocks). For the second phase of the experiment, block sizes of 30 and 50 both performed well with a 0\% FAR. However, a block size of 30 had a 5\% FRR with 1.82 average blocks to detection and a block size of 50 had a 4\% FRR with 2.28 average blocks to detection. A block size of 30 is preferable because a small increase in FRR is worth a smaller block size and quicker detection. Problems caused by a small increase in the FRR, meaning slightly more genuine users are being detected as intruders, can be solved by requiring that the genuine user re-authenticate to access the system. However, the average number of blocks that it takes to detect the user as genuine or an intruder reduces to 1.87 with 30 characters per block, which is comparatively better because the intruder only has about 56 keystrokes to gain root access in a system. This is better than 2.28 blocks with 50 characters per block giving an intruder about 114 keystrokes to compromise a system.

This experimentation was carried out on a one-on-one comparison basis where each user was tested against all the others, including themselves. Some results showed high correlation between two users typing feature sets and that indicates why some users were wrongly classified. In other words, not surprisingly, users that type more similarly to each other are more likely to be wrongly classified when tested against each other's models. This could explain why the second phase of the initial experiment performed better than the first. Recall that the first phase gave each volunteer a prompt and a window to type in and the second phase gave each volunteer free use of their computer. The variations between predictions between the two halves of the experiment, where the freestyle computer usage outperforms the prompted usage, could be explained by taking into account user's behavior. When given a certain topic to type on, users are more likely to type in similar patterns. However, when users can use their computer however they want (complete assignments, send out emails, chat on social media), they are more likely to type differently as the model captures their true self and their true behavior, which is more unique to each individual.

After determining the initial success of Spy Hunter, the extended experimentation provided validation of the results. This showed that the data in the original experimentation had not been over-fit and that the obtained results were legitimate. Further, the results in the extended experimentation demonstrated little variation between segmenting the data into 5 data sets versus 10 sets. The average number of blocks varied between 2.6813 and 2.9834 blocks. Since each block is 80 characters, these two results vary by only 24 characters (0.3021 blocks). In the worst case, the prompted data set with 10 folds, less than 240 characters are required to validate a user. Further, the second freestyle phase of data collection outperformed the first prompted phase in the extended experimentation. The second phase held a lower FAR and average number of necessary blocks, regardless of the number of folds. The only metric that the first phase surpassed the second phase was FRR and only by .35\% in the 5 fold experiments. Because a genuine user can re-authenticate to gain back access after being falsely rejected, this metric hold less weight than FAR, since a false acceptance is detremental to the security of the system. 

\subsection{Limitations}
Data was collected from users remotely. Users were in their own environment to allow for simulation of "normal activity". However, limitations exist on capturing natural activity because volunteers knew their keystrokes were being recorded. Because they knew that their typing was being analyzed, they likely didn't act truly naturally. Additionally, users participated in different activities during the free usage data collection phase. Someone who was playing computer games while being recorded would likely be validated correctly against someone who was shopping online while being recorded. Although this leads to better results, it isn't indicative of long term authentication since behaviors change over time. It's possible that both participants play video games and shop online, but only happened to be doing different activities during data collection. This underscores the need for truly continuous authentication.

Additionally, our initial experimentation approach utilized a simplified K-fold cross-validation method. This allowed us to gather promising results to quickly gauge the implications of Spy Hunter, however they required verification by the extended experimentation. Without the extended experimentation, the initial experimentation doesn't provide enough evidence of Spy Hunter's successful authentication. 

\subsection{Future Research}
Continuous authentication systems still create vulnerabilities in the system. By collection personal user data and storing such data on a given device, we create a target for adversaries and a potential vulnerability. Some researchers have targeted this vulnerability and sanitization schemes to detect and remove personal data from the collected data \cite{Sun_2015}. Since Spy Hunter only relies on the timing of keypresses and not which key is being pressed, it allows for anonymous collected data. Provided Spy Hunter's continued success, user privacy motivates the addition of further sanitization techniques to the underlying mechanism. This would provide another layer of protection to the user without any additional burden on the user. 

Also, Spy Hunter authenticates based on blocks of data at a time. Successful periodic authentication indicates that Spy Hunter can be extended to fully continuous authentication where the data input is a constant flow of keypresses rather than blocks of keypresses. This presents computation efficiency issues since it is harder to implement Spy Hunter on each individual key press than it is to implement it with segments of data input. A successful authentication mechanism can't be bogged down with computation, otherwise reaction time can be slowed and give intruders more opportunity to attack. Future implementations of Spy Hunter will ensure continuous authentication without being slowed. 

\section{Conclusion}
Spy Hunter, a continuous authentication mechanism uses keystroke dynamics to validate users over blocks of data. Continuous authentication ouperforms one-time verification systems. Utilizing K-fold cross-validation techniques, Spy Hunter is assessed based on FAR, FRR, and the number of required blocks of data. Spy Hunter maintains error rates below 6\% and identifies users in minimal numbers of keystrokes. Technological advancement necessitates reliable authentication mechanisms, especially across fully-incorporated systems. These systems currently rely on a moment of authentication, however continuous authentication systems determine a users identity utilizing a constant biometric analysis. Spy Hunter specifically focuses on the timing of different keystrokes rather than the actual key being pressed. This allows for anonymous data to authenticate users. This easily-incorporated periodic biometric authentication system validates genuine users and efficiently detects intruders. Because it verifies users behind the scenes, Spy Hunter is not constrained to a particular process or machine. Instead, it is flexible and can be layered with other mechanisms to provide high-level security. Where other continuous authentication techniques rely on scripted typing, Spy Hunter validates over free text in authentic environments. This is accomplished in two phases, one where the user is provided a prompt and another where the user is allowed free access to their computer. Spy Hunter expands on the possibilities for behavioral analysis based on keystroke dynamics. The end results show promise for continued research in Spy Hunter and indicates future success in continuous authentication. 

\bibliographystyle{ACM-Reference-Format}
\bibliography{refs.bbl}

\end{document}